\documentclass[fleqn]{2020SCGE}
\usepackage{epstopdf}
\usepackage{cite}
\usepackage{graphics}

\begin{document}
%\captionsetup{singlelinecheck=off}

\ensubject{subject}

%%%%%%%%%%%%%%%%%%%%%%%%%%%%%%%%%%%%%%%%%%%%%%%%%%%%%%%
%%% Authors do not modify the information below
%%% ????????????????
%%% ??????????, ????????????{}, ???????????????????
%Letter to the Editor??Article%??????
\ArticleType{Article}%??Article
\SpecialTopic{SPECIAL TOPIC: The Chinese H$\alpha$ Solar Explorer --- CHASE Mission}%???????
\Year{2022}
\Month{March}
\Vol{65}
\No{??}
\DOI{10.1007/s11433-022-1893-3}
\ArtNo{289602}
\ReceiveDate{February 27, 2022}
\AcceptDate{March 16, 2022}
%\OnlineDate{January 1, 2016}
%%%%%%%%%%%%%%%%%%%%%%%%%%%%%%%%%%%%%%%%%%%%%%%%%%%%%%%

%%% title: ????
%%% \title{title}{title for citation}
\title{The Chinese H$\alpha$ Solar Explorer (CHASE) mission: An overview}{The Chinese H$\alpha$ Solar Explorer (CHASE) Mission: An overview}

%%% Corresponding author: ???????
%%%   \author[number]{Full name}{{email@xxx.com}}
%%% General author: ???????
%%%   \author[number]{Full name}{}
\author[1,2]{Chuan Li}{{lic@nju.edu.cn}}
\author[1,2]{Cheng Fang}{{fangc@nju.edu.cn}}
\author[1,2]{Zhen Li}{}
\author[1,2]{MingDe Ding}{}
\author[1,2]{PengFei Chen}{}
\author[1,2]{Ye Qiu}{}
\author[3]{\\Wei You}{}
\author[3]{Yuan Yuan}{}
\author[3]{MinJie An}{}
\author[4]{HongJiang Tao}{}
\author[4]{XianSheng Li}{}
\author[4]{Zhe Chen}{}
\author[4]{Qiang Liu}{}
\author[4]{\\Gui Mei}{}
\author[4]{Liang Yang}{}
\author[5]{Wei Zhang}{}
\author[5]{WeiQiang Cheng}{}
\author[5]{JianXin Chen}{}
\author[3]{ChangYa Chen}{}
\author[3]{\\Qiang Gu}{}
\author[3]{QingLong Huang}{}
\author[3]{MingXing Liu}{}
\author[4]{ChengShan Han}{}
\author[4]{HongWei Xin}{}
\author[4]{\\ChangZheng Chen}{}
\author[1,2]{YiWei Ni}{}
\author[1,2]{WenBo Wang}{}
\author[1,2]{ShiHao Rao}{}
\author[1,2]{HaiTang Li}{}
\author[3]{Xi Lu}{}
\author[3]{\\Wei Wang}{}
\author[6]{Jun Lin}{}
\author[7]{YiXian Jiang}{}
\author[8]{LingJie Meng}{}
\author[8]{Jian Zhao}{}

%%% Author information for page head. ?¨¹?§Ö????????
%%% ??????????????, ??????????author???
\AuthorMark{C. Li, C. Fang, Z. Li, M. D. Ding, P. F. Chen, Y. Qiu}%\authorcr????????

%%% Authors for citation. ????????§Ö????????
%%% ??????????????, ??????????author???
\AuthorCitation{C. Li, C. Fang, Z. Li, M. D. Ding, P. F. Chen, Y. Qiu, et al.}

%%% Address. ???
%%%   \address[number]{Address, City {\rm Postcode}, Country}
\address[1]{School of Astronomy and Space Science, Nanjing University, Nanjing 210023, China}
\address[2]{Key Laboratory for Modern Astronomy and Astrophysics, Ministry of Education, Nanjing 210023, China}
\address[3]{Shanghai Institute of Satellite Engineering, Shanghai 201109, China}
\address[4]{Changchun Institute of Optics, Fine Mechanics and Physics, University of Chinese Academy of Sciences, Changchun 130033, China}
\address[5]{Shanghai Academy of Spaceflight Technology, Shanghai 201109, China}
\address[6]{China Center for Resources Satellite Data and Application, Beijing 100094, China}
\address[7]{China Aerospace Science and Technology Corporation, Beijing 100048, China}
\address[8]{Key and Special Project Center of China National Space Administration (CNSA), Beijing 100101, China}

%\contributions{}%????????

%%% Abstract. ??
\abstract{The Chinese H$\alpha$ Solar Explorer (CHASE), dubbed ``Xihe" --- Goddess of the Sun, was launched on October 14, 2021 as the first solar space mission of China National Space Administration (CNSA). The CHASE mission is designed to test a newly developed satellite platform and to acquire the spectroscopic observations in the H$\alpha$ waveband. The H$\alpha$ Imaging Spectrograph (HIS) is the scientific payload of the CHASE satellite. It consists of two observational modes: raster scanning mode and continuum imaging mode. The raster scanning mode obtains full-Sun or region-of-interest spectral images from 6559.7 to 6565.9 \AA{} and from 6567.8 to 6570.6 \AA{} with 0.024 \AA{} pixel spectral resolution and 1 minute temporal resolution. The continuum imaging mode obtains photospheric images in continuum around 6689 \AA{} with the full width at half maximum of 13.4 \AA{}. The CHASE mission will advance our understanding of the dynamics of solar activity in the photosphere and chromosphere. In this paper, we present an overview of the CHASE mission including the scientific objectives, HIS instrument overview, data calibration flow, and first results of on-orbit observations.}

%%% Keywords. ?????
\keywords{Space-based telescope, Solar physics, Chromosphere, Photosphere}
\PACS{95.55.Fw, 96.60.--j, 96.60.Na, 96.60.Mz}

\maketitle

%\tableofcontents%?????
%%%%%%%%%%%%%%%%%%%%%%%%%%%%%%%%%%%%%%%%%%%%%%%%%%%%%%%
%%% The main text. ???????
%???????????????????\cref{fig1}
%\twocolumn\onecolumn
%%%%%%%%%%%%%%%%%%%%%%%%%%%%%%%%%%%%%%%%%%%%%%%%%%%%%%%

\begin{multicols}{2}	
\section{Introduction}\label{section1}

\begin{figure*}[t]
	\centerline{\includegraphics[width=0.98 \textwidth]{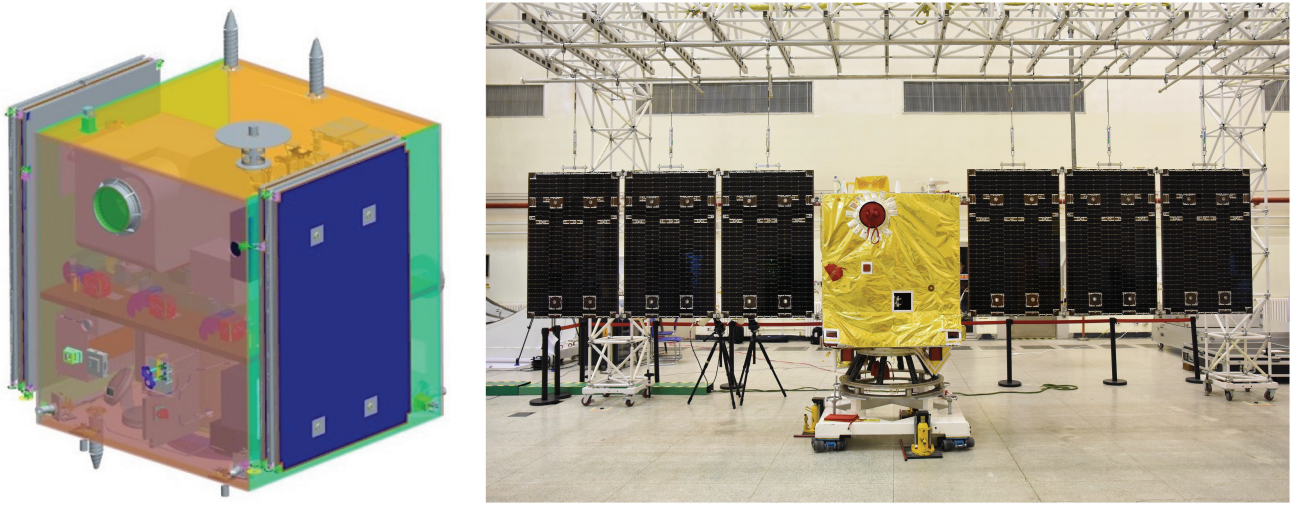}}
	\caption{The schematic view of the CHASE satellite (left panel) and the photograph at the stage of assembling (right panel).}
	\label{satellite}
\end{figure*}

The H$\alpha$ line is one of the most important optical lines for solar observations. The spectral profile near the H$\alpha$ line center at 6562.8 \AA{} reflects the information of the solar chromosphere, and the far wings reflect the information of the photosphere. Hence, the whole spectral profile can be used to decipher the detailed structure of the solar lower atmosphere. The very first solar H$\alpha$ images were obtained by George E. Hale on Mount Wilson Solar Observatory \cite{hale}. Since then H$\alpha$ spectroscopic observations have been carried out in ground-based telescopes \cite{fang1,liu,potzi,fang2,rao1,yan}. One may notice that there were very few full-Sun H$\alpha$ spectroscopic observations except the Solar Dynamics Doppler Imager (SDDI) installed at Hida observatory, Japan \cite{ichimoto}, and the spectroheliograph patrol telescope at Kislovodsk mountain astronomical station, Russia \cite{berezin}. The SDDI uses a tunable filter to obtain H$\alpha$ images with a passband of 0.25 \AA{}. The spectroheliograph of Kislovodsk station has a spectral resolution of 0.16 \AA{}. However, due to the seeing effects arising from the Earth's atmosphere, the designed spatial resolution for the ground-based telescopes cannot be easily reached except using the adaptive optical systems \cite{rao2}. Another limitation for the ground-based telescopes is that they cannot achieve all-day and all-weather solar observations. It is therefore necessary to promote a solar space mission to carry out the full-Sun H$\alpha$ spectroscopic observations with high temporal and spectral resolutions.

The Chinese H$\alpha$ Solar Explorer (CHASE) satellite, with a weight of 508 kg and a size of 1210 mm $\times$ 1210 mm $\times$ 1350 mm, was designed to test a newly developed satellite platform and to perform spectroscopic solar observations in the H$\alpha$ waveband. The CHASE satellite was launched into a Sun-synchronous orbit using a CZ-2D rocket at 18:51:04 on October 14, 2021 (China Standard Time, CST). The orbit has an average altitude of $\sim$517 km and a period of $\sim$95 minutes. The lifetime of the CHASE mission is designed to be 3 years, which fortunatelly covers the ascending phase till the maximum of Solar Cycle 25. Along with the X-ray and Extreme Ultraviolet Imager (X-EUVI) onboard the FY-3E mission launched on July 5, 2021 \cite{zhangp} and the Advanced Space-based Solar Observatory (ASO-S) scheduled to be launched in the last quarter of 2022 \cite{gan}, the CHASE mission marks a milestone for the Chinese solar physics to glide into the space age \cite{chen1}.

Figure 1 shows the schematic view of the CHASE satellite (left panel) and its photograph prior to launch (right panel). The ultra-precision satellite platform of the CHASE mission was designed with a pointing accuracy of $5 \times 10^{-4}$$^{\circ}$ and a stability of $5 \times 10^{-5}$$^{\circ} / \rm s$. Detailed designment and on-orbit performance of the satellite platform are described in an accompanied paper by Zhang et al. \cite{zhang1}. The excellent pointing accuracy and stability of the satellite platform ensures solar spectroscopic observations for the scientific payload --- H$\alpha$ Imaging Spectrograph (HIS), without requirement for a guide telescope or an imaging stabilization system.

The CHASE mission was dubbed ``Xihe" in Chinese --- Goddess of the Sun. A brief introduction of the CHASE mission was previously reviewed by Li et al. \cite{lic}. There have been many improvements in the satellite and scientific payload since then. This paper presents the scientific objectives of the CHASE mission, the HIS instrument overview and its techonical parameters, data calibration flow, and the first results of on-orbit observations.

\section{Scientific objectives}\label{sec:2}

The primary goal of the CHASE mission is to investigate the dynamics of solar activity in the lower atmosphere, namely the photosphere and the chromosphere, and to understand the physical mechanisms of solar eruptions. The full-Sun spectroscopic observations are also useful in Sun-as-a-star studies that provide implications to stellar physics. We discuss here some of the main scientific objectives to be studied by the CHASE mission.

\begin{figure*}[t]
	\centerline{\includegraphics[width=0.92 \textwidth]{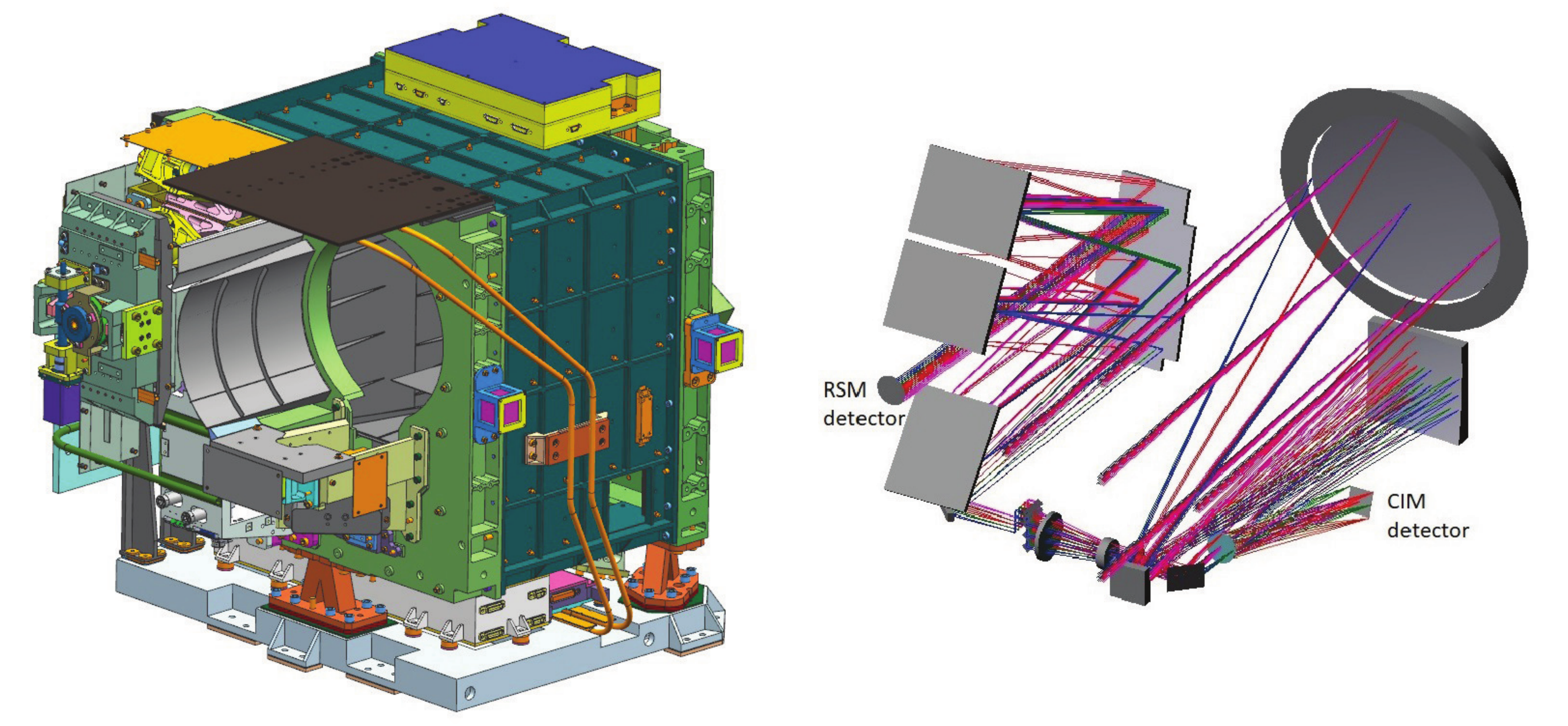}}
	\caption{The schematic view of the HIS instrument (left panel) and its optical design illustrating two observational modes of RSM and CIM (right panel).}
	\label{instrument}
\end{figure*}

\subsection{Formation, dynamics, and chirality of solar filaments}

The H$\alpha$ line is the strongest spectral line in the solar visible spectrum, and is one of the best spectral lines for filament or prominence observations. A filament is a volume of cool material, compared to the surrounding hot plasmas, situated above the polarity inversion line. There are many controversial topics related to solar filaments, e.g., how they are formed, how the longitudinal and transverse oscillations can be utilized to diagnose the elusive coronal magnetic field, and how they are triggered to erupt \cite{chen2}. With H$\alpha$ spectroscopic observations, we can derive the 3-dimensional velocities of the filament counterstreamings \cite{zhou}, investigate the filament interaction and oscillations that provide important clues for solar eruptions \cite{chen3,yang}. Solar filaments are supported by either sheared magnetic arcades \cite{devore} or magnetic flux ropes \cite{cheng}, both of which are believed to be the pre-eruption magnetic structures of coronal mass ejections \cite{ouyang}. Using the H$\alpha$ spectroscopic observations of the CHASE mission, one can derive the filament topology and chirality that are useful to diagnose the supporting magnetic fields \cite{chen4,casini}.

\subsection{Dynamics of solar activity in photosphere and chromosphere}

The CHASE mission provides solar spectroscopic observations with high spectral and temporal resolutions at the wavebands of H$\alpha$ (6562.8 \AA{}), Fe I (6569.2 \AA{}), and Si I (6560.6 \AA{}), which will be described in Section \ref{sec:4}. Hereafter the wavelength refers to the air wavelength in comparison with the ground-based observations. The H$\alpha$ line is the strongest chromospheric line, and the other two are photospheric lines. Therefore the dynamics of solar activity in both the photosphere and the chromosphere can be revealed by the CHASE mission simultaneously. How is the magnetic energy released and transported in white-light flares \cite{ding} and Ellerman bombs \cite{fang3}, via thermal or non-thermal energy dissipation? What are the precursors and triggering mechanisms of solar eruptions \cite{wang}? What is the difference between the differential rotations of the photosphere and the chromosphere \cite{schroter,lik}? How are materials tranported from the lower atmosphere to the corona \cite{brooks}? These questions can be anwsered by the full-Sun spectroscopic observations of the CHASE mission itself, or by combing spectral and imaging observations from other solar missions, e.g., Solar Dynamics Observatory (SDO) \cite{pesnell}, Interface Region Imaging Spectrograph (IRIS) \cite{pontieu}, Solar Orbiter (SolO) \cite{muller}, and ASO-S \cite{gan}.

\subsection{Comparative studies of solar and stellar magnetic activities}

Stellar magnetic activities are now becoming hot topics in astrophysics. Stellar flares, many of which are ``superflares", have been observed and studied on a variety of stars \cite{schaefer,balona}. Similar to the magnetic energy release on the Sun, stellar flares are believed to be closely associated with stellar filament eruptions and CMEs, which may play an important role in mass and momentum loss of stars, and in producing hazard space weather of the exoplanetary environment. So far stellar CMEs have not been well observationally explored \cite{odert,argiroffi}. The high temporal and spectral resolutions of the CHASE mission facilitate the study of the evolution of integrated H$\alpha$ profiles during solar eruptions, which can shed light on the understanding of similar processes on Sun-like stars. That is to say, the full-Sun H$\alpha$ spectroscopic observations provide a unique tool to study the Sun as a whole and to compare the spectral differences, e.g., Balmer line asymmetries, between stellar eruptions and solar eruptions.

\begin{table}[H]
	\footnotesize
	\caption{The HIS instrument characteristics.}
	\label{table1}
	\begin{tabular}{lll}     % define the column alignment % l: left, c: center, r: right
		\hline                   % horizontal line
	    Systems & Items & Values  \\
		\hline
    	&	Mass & 54.9 kg  \\
        Mechanics & Size & 635 mm $\times$ 556 mm $\times$ 582 mm  \\
	    &	Power & Average: 58 W, Maximum: 98 W  \\
		\hline
    	&	Primary aperture & 180 mm      \\
     	Optics & Effective focal length & 1820 mm      \\
	    & Field of view &  40$^{'}$ $\times$ 40$^{'}$     \\		
		\hline
        RSM 	& Array	 &   4608 $\times$ 376 (window applied)     \\
        detector	&	 Pixel size   &   4.6  $\mu$m    \\
     	&	 Quantization (ADC)   &    12 bit    \\
    	&	  Full well   &    14.5 k    \\
        CIM	&	Array  &     5120 $\times$ 5120     \\
        detector	&	  Pixel size   &    4.5  $\mu$m    \\
     	&	  Quantization (ADC)   &    10 bit    \\
     	&	  Full well   &    12 k    \\		
		\hline
     	&	  Transmission rate   &    300 Mbps    \\
        Telemetry	&	  Ground capture   &    $\sim$1.2 Tb per day (compressed)   \\
    	&	  Data compression   &    6:1 (typical)   \\				
		\hline
	\end{tabular}
\end{table}

\section{Instrument overview}\label{sec:3}

There are two payloads onboarded the CHASE satellite. The solar atomic frequency discrimanitor aims to test its accuracy of spectral velocity measurement and experiment the potential ability of autonomous navigation for future space explorations \cite{Zhang2}. The HIS instrument is the scientific payload of the CHASE mission. It utilizes the excellent pointing accuracy and stability of the satellite platform to provide solar spectroscopic observations in the H$\alpha$ waveband. The HIS instrument has a weight of 54.9 kg and a size of 635 mm $\times$ 556 mm $\times$ 582 mm. It consists of three optical systems: preposition optical system, raster scanning system, and continuum imaging system. The schematic view of the HIS instrument and its optical systems are shown in Figure 2. Table 1 summarizes the characteristics of the HIS instrument. Detailed instrument designment and on-orbit performance are described in an accompanied paper by Liu et al. \cite{Liu1}.

The preposition optical system of HIS instrument is composed of a filter assembly and an off-axis three-mirror-anastigmatic (TMA) assembly. The filter assembly consists of two radiation-shielding glasses that are coated with a bandpass film and an infrared-cut film, respectively, allowing the transmission of the passband of 6430 -- 6830 \AA{}. The TMA assembly consists of three reflecting mirrors and possesses characteristics of wide Field of View (FoV) and long focal length in a limited volume of optical system \cite{chenz}. Specifically, the HIS instrument has a focal length of 1820 mm and an effective aperture of 180 mm, leading to an F-number of 10.1 and a FoV of $40^{'} \times 40^{'}$.

The raster scanning system is composed of a slit, a folding mirror, a plane grating, a collimating mirror, an imaging mirror, and a CMOS detector. The length of the slit is 23 mm, and the width is 9 $\mu$m. The grating groove density is 1900 lp/mm. The pixel size of the CMOS detector is 4.6 $\mu$m. These lead to the instrument spectral resolution, in other words, the spectral full width at half maximum (FWHM) of 0.072 \AA{}, and the pixel spectral resolution of 0.024 \AA{}. Another key component of the raster scanning system is a scanning unit that moves the solar imaging disk across the slit with a constant speed. It is driven by a linear electric motor, whose experimental scanning speed is 4.6 $\pm$ 0.3 mm/s. To complete a full-Sun scanning, only $\sim$46 seconds is needed. A redundancy of 1-minute scanning is designed to improve the system reliability. For a region-of-interest scanning, a temporal resolution of 30 -- 60 seconds can be achieved.

The continuum imaging system is composed of a beam splitter, a neutral density filter, a narrow bandpass filter, and a CMOS detector. The neutral density filter aims to attenuate the solar continuum emission, with a transmittance rate of 1/5000. The narrow bandpass filter has a FWHM of 13.4 \AA{} with the centeral wavelength at 6689 \AA{}, which is located in a pure continuum waveband and is not contaminated by any spectral lines. The CMOS detector has 5120 $\times$ 5120 pixels, and the pixel size is 4.5 $\mu$m. These lead to a pixel spatial resolution of 0.52 arcsec.

\begin{figure*}[t]
	\centerline{\includegraphics[width=0.95 \textwidth]{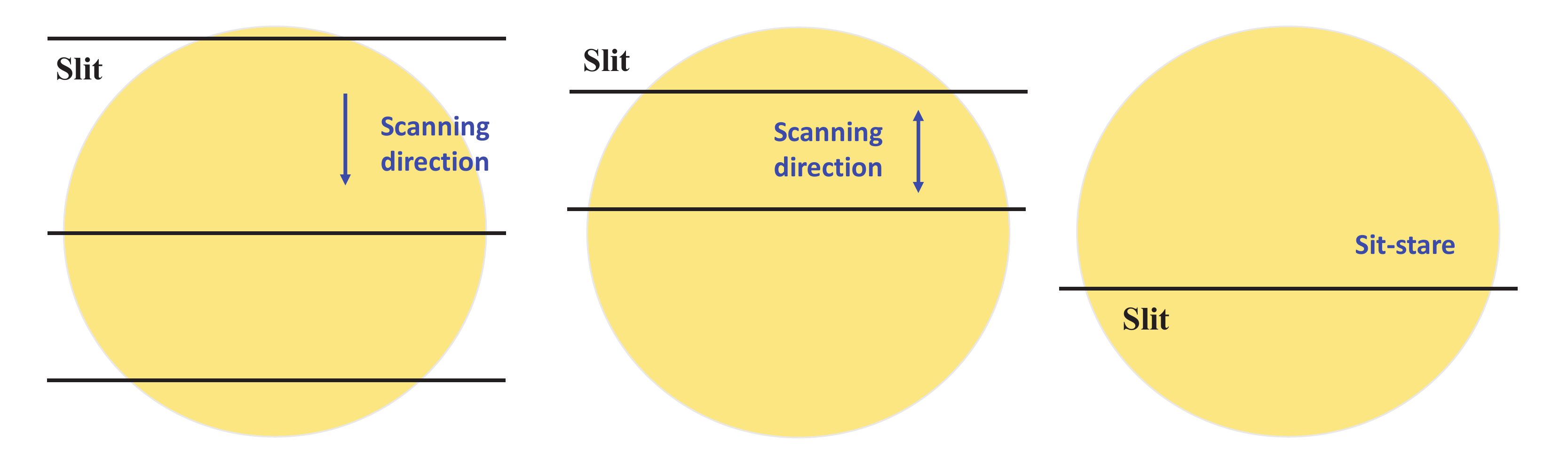}}
	\caption{The three sub-modes of the RSM observational mode. From left to right: the full-Sun scanning, region-of-interest scanning, and sit-stare spectroscopy. Note that the solar imaging disk, rather than the slit, is moving during scanning.}
	\label{observationmode}
\end{figure*}

The HIS instrument can operate in two observational modes: raster scanning mode (RSM) and continuum imaging mode (CIM). The RSM mode acquires solar spectra in wavebands of H$\alpha$ (6559.7 -- 6565.9 \AA{}) and Fe I (6567.8 -- 6570.6 \AA{}) with high spectral and temporal resolutions. The RSM mode has three sub-modes, namely the full-Sun scanning, region-of-interest scanning, and sit-stare spectroscopy. The schematic sketch in Figure 3 shows the relative positions of the slit against the solar imaging disk. It should be mentioned that the solar imaging disk, rather than the slit, is moving during scanning. The full-Sun and region-of-interest scanning have temporal resolutions of 30 -- 60 seconds. The sit-stare spectrocopy has a temporal resolution of $<$ 10 ms. The CIM mode is designed to obtain the high-cadence (1 second) full-Sun photospheric images that can be applied to verify the stability of the satellite platform. Table 2 summarizes the parameters of the two observational modes.

\begin{figure*}[t]
	\centerline{\includegraphics[width=0.95 \textwidth]{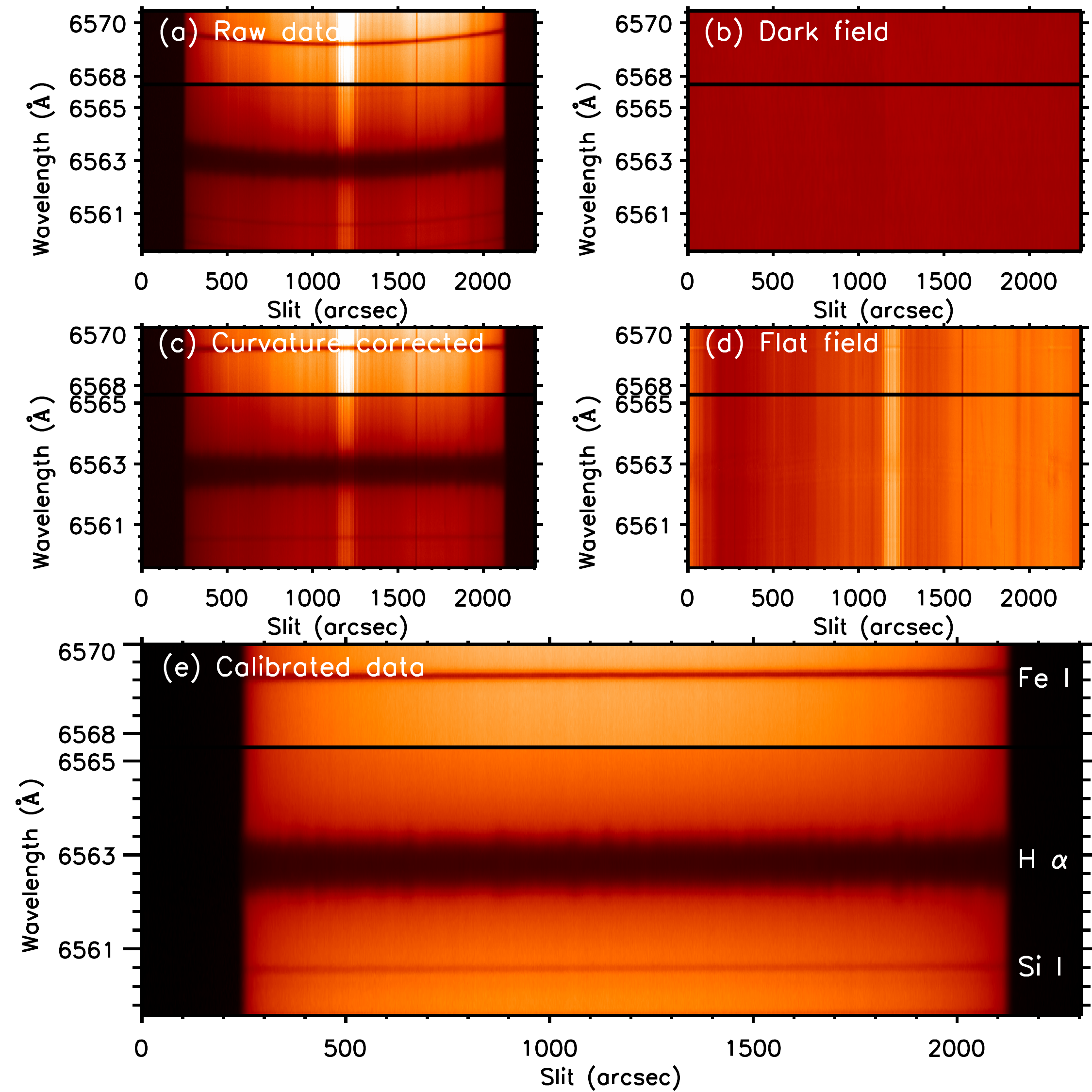}}
	\caption{The calibration flow of the RSM spectra from Level 0 raw data to Level 1 science data. Two passbands of 6559.7 -- 6565.9 \AA{} and 6567.8 -- 6570.6 \AA{} are displayed, which are seperated by the black line. The spectra were obtained at 00:53:14 UT on October 24, 2021 when the slit was located at the center of the solar disk. (a) The Level 0 raw spectra show not only the spectral information, but also the systematic artifacts, digital offsets, slit image curvature, etc. (b) The dark field was obtained when the CHASE/HIS is pointing to the dark cold space. (c) The slit image curvature is corrected by using experimental curvature coefficients. (d) The caluclated flat field shows the systematic vignetting, artifacts on the slit and detector, etc. (e) The calibrated Level 1 spectra show clearly the spectral lines of H$\alpha$, Fe I, and Si I that can be used for science purposes.}
	\label{datacalibration}
\end{figure*}

\begin{figure*}[t]
	\centerline{\includegraphics[width=0.95 \textwidth]{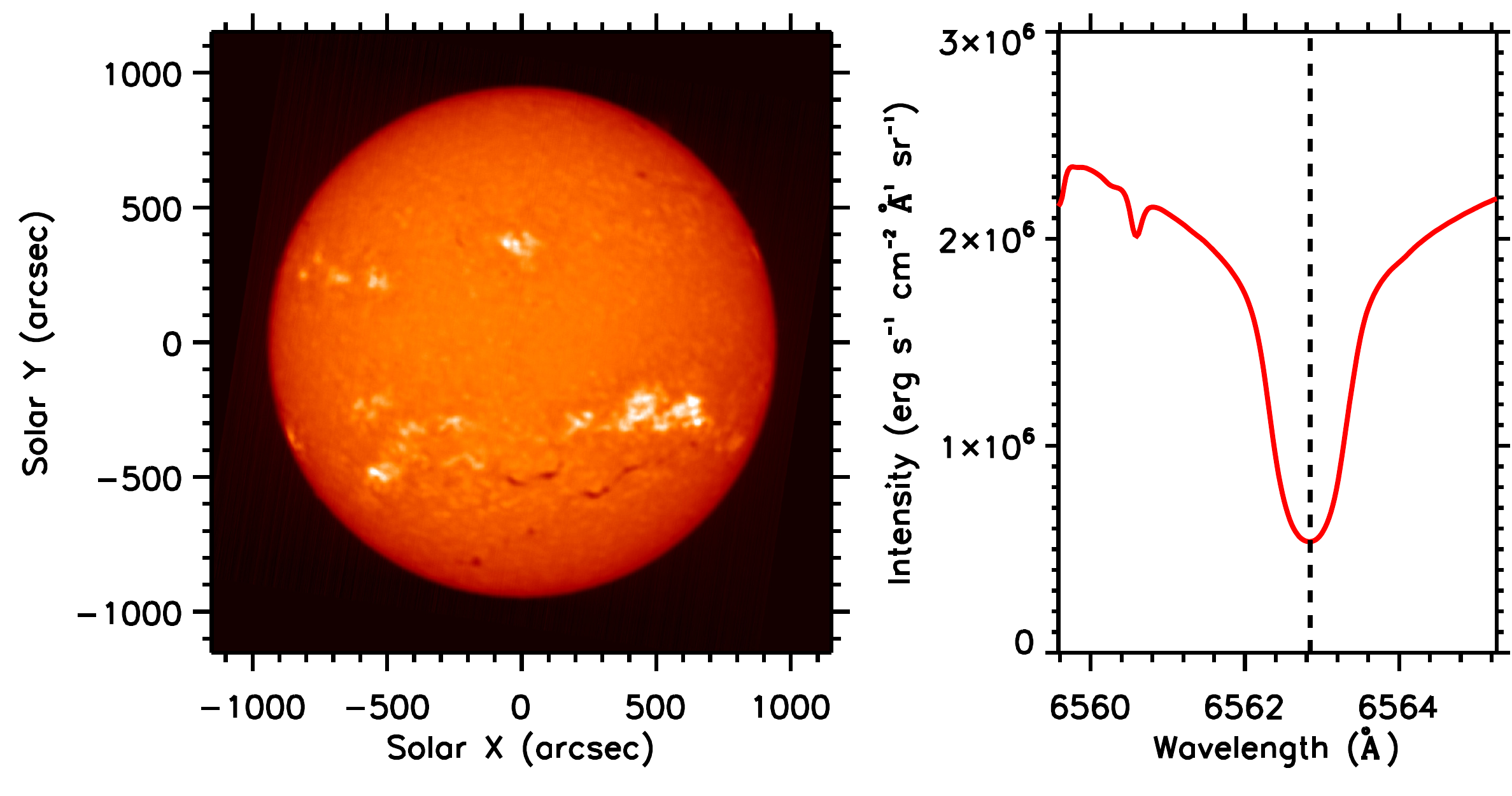}}
	\caption{Left panel: the spectroscopic image of the full solar disk near the H$\alpha$ center at 6562.8 \AA{}. The scanning time was from 06:01:05 -- 06:01:52 UT on December 22, 2021. Right panel: the disk-center averaged H$\alpha$ line profile with a pixel spectral resolution of 0.024 \AA{}. The Si I line centered at 6560.6 \AA{} is located at the H$\alpha$ blue wing.}
	\label{haimage}
\end{figure*}

\begin{table}[H]
	\footnotesize
	\caption{The parameters of the HIS observational modes}
	\label{table2}
	\begin{tabular}{lll}     % define the column alignment % l: left, c: center, r: right
		\hline                   % horizontal line
    Modes	& Items & Values  \\
		\hline
	&	Passbands & H$\alpha$: 6559.7 -- 6565.9 \AA{}  \\
	&	& Fe I: 6567.8 -- 6570.6 \AA{}  \\
	&	Instrument FWHM &  0.072 \AA{} \\
	RSM &	Pixel spectral resolution & 0.024 \AA{}  \\
	&	Pixel spatial resolution & 0.52$^{''}$  \\
	&	Full-Sun scanning time & 60 seconds  \\
	&	Local-area scanning time & 30 -- 60 seconds  \\
	&	Exposure time for each step  &  $<$ 10 ms  \\		
		\hline
	&	Center wavelength &     6689 \AA{}   \\
	&	Passband FWHM &      13.4 \AA{}    \\
	CIM &	Pixel spatial resolution & 0.52$^{''}$  \\
	&	Exposure time & $<$ 5 ms  \\
	&	Frame rate &  1 fps \\
		\hline
	\end{tabular}
\end{table}

\section{Data processing}\label{sec:4}

After the successful launch of the CHASE mission on October 14, 2021 (CST), the HIS instrument saw its first light on October 24, 2021. The CHASE mission does not produce science data in orbit, but transmits the raw data to three ground stations (Miyun, Kashi, and Sanya) located in China. The raw data are then transfered through a dedicated internet access to the Solar Science Data Center of Nanjing University (SSDC-NJU), where the Level 0 data are produced, archived, and further calibrated to the Level 1 data and higher-level products. Here a brief calibration flow is introduced based on the first-light observations. Detailed calibration procedures are described in an accompanied paper by Qiu et al. \cite{Qiu1}.

\subsection{Calibration flow}

According to the transmission rate of the downlink stations and the typical satellite transit time, the ground capture of the raw data is $\sim$1.2 Tb per day. The raw data are JPEG2000 compressed and stored at SSDC-NJU. They are then converted into the Level 0 data that are Rice compressed to reduce storage requirement. The Level 0 image files are archived as 4608 $\times$ 376 arrays for the RSM mode and 5120 $\times$ 5120 arrays for the CIM mode. Taking the full-Sun RSM observation as an example, one sequence of scanning generates $\sim$14.9 GB decompressed Level 0 data. Therefore a powerful pipeline system is required for the data acquisition and processing. The computing system at SSDC-NJU has the storage capacity of $\sim$6 Pb and the computing ability of $\sim$102.4 Tflops.

Processing Level 0 data to Level 1 science data involves several steps. Here we focus on the calibration of the RSM spectra. At first, the dark field including digital offsets, read noise, and dark current of the detector, is removed. Step 2 is the correction of the slit image curvature, which is a common phenomenon in off-axis-mirror spectrometers and has been attributed to both the off-axis mirrors and the diffraction plane grating. The correction can be carried out by using the experimental curvature coefficients, which are calculated by the pixel positions of the wavelength tunable laser. Step 3 is the correction of the flat field. It aims to remove the systematic vignetting, the artifacts on the slit and detector, and the intensity patterns due to irregularities of the slit width. To obtain the flat field, the center of the solar imaging disk should be moving along the slit, during which the spectra are simultaneously recorded. These center-position spectra do not contain information of solar limb darkening and solar differential rotation, and can be used to derive the flat field. After the above procedures, along with the  coordinates transformation, wavelength and intensity calibrations, the RSM Level 1 science data are produced. For more details, one may refer to Qiu et al. \cite{Qiu1}. Figure 4 shows how the RSM Level 0 data are processed into the calibrated science data based on the first-light spectra observed on 24 October, 2021.

The RSM Level 1 data are Rice compressed and seperately archived as 4608 $\times$ 260 arrays for the H$\alpha$ waveband (6559.7 -- 6565.9 \AA{}) and 4608 $\times$ 116 arrays for the Fe I waveband (6567.8 -- 6570.6 \AA{}), respectively. Here the number of 4608 refers to the pixels along the slit, whereas 260 and 116 refer to the pixels of the wavelength. For science users, we usually provide 3-dimensional arrays with the third dimension referring to the steps of a scanning sequence. Taking one specific wavelength along the slit from every step, we can splice a solar image at this wavelength. Therefore one scanning sequence produces 376 solar images at different wavelengths. Figure 5 shows an example of the full-Sun spectroscopic image at the H$\alpha$ center of 6562.8 \AA{}. The full-Sun scanning time period was from 00:52:49 -- 00:53:35 UT on 24 October 2021. Note that the spectral line located in the H$\alpha$ blue wing is the Si I line centered at 6560.6 \AA{}.

\begin{figure}[H]
	\centerline{\includegraphics[width=0.48 \textwidth]{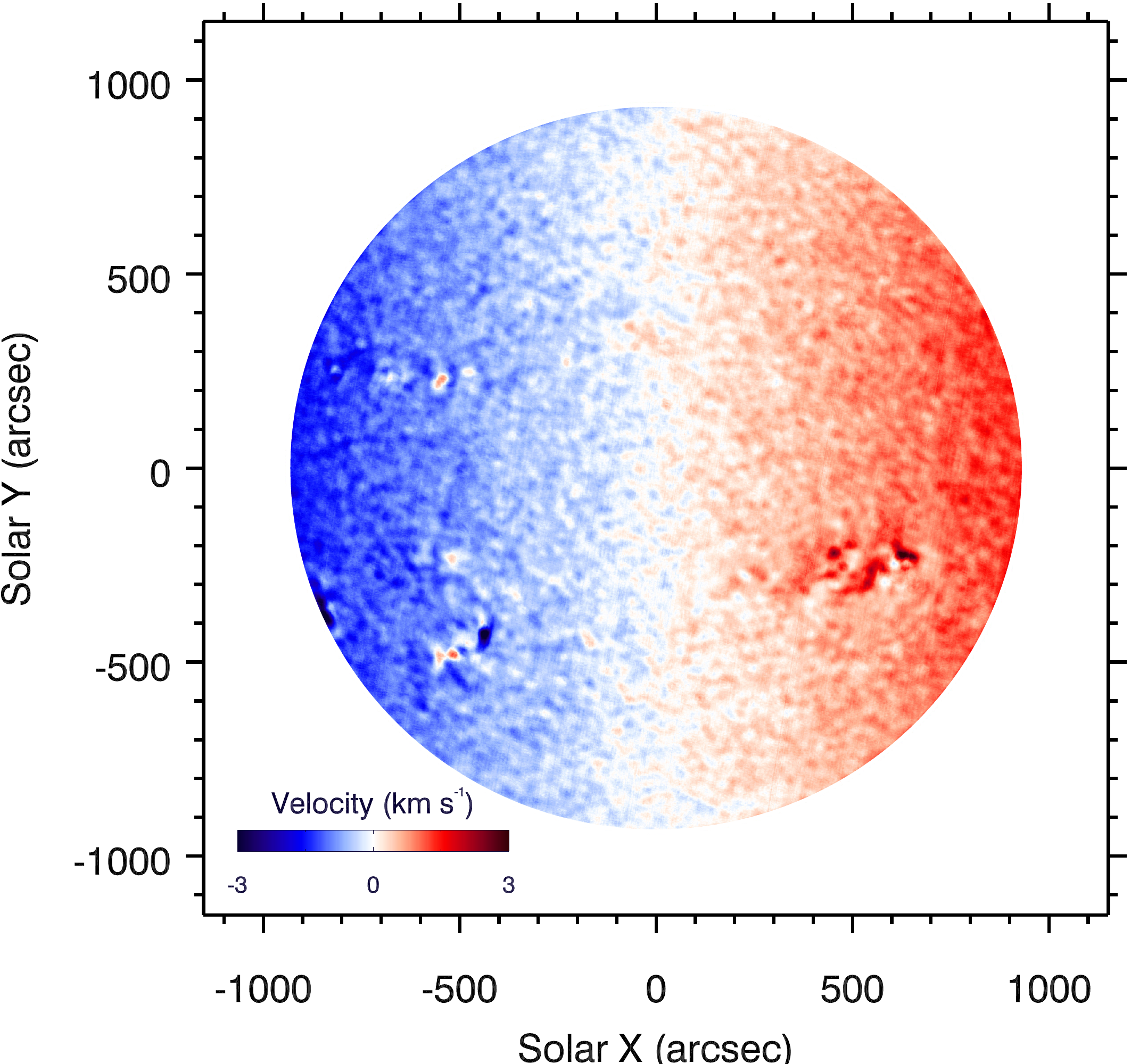}}
	\caption{The full-Sun chromospheric Dopplergram at 06:01:05 -- 06:01:52 UT on December 22, 2021 derived from the H$\alpha$ profiles by using the cross-correlation method.}
	\label{dopplergram}
\end{figure}

The RSM Level 1 data can be further applied to produce higher-level products. For instance, the absolute intensity calibration can be done based on the spectral profiles and the emission measure of the continuum spectra near the H$\alpha$ waveband. Other important products include the full-Sun or region-of-interest photospheric and chromospheric Dopplergrams, which can be derived simutaneously in less than 1 minutes from the profiles of the H$\alpha$, Fe I or Si I spectral lines. Figure 6 shows the full-Sun chromospheric Dopplergram that is derived from the full-Sun H$\alpha$ profiles by using the cross-correlation method. Specifically, the Doppler velocities are calculated by comparing the spectral profile of each pixel with a reference profile, which is an average profile over a central region of the solar disk. The accuracy of the velocity field is calculated to be $\sim$0.06 $\rm km \ s^{-1}$. The differential rotation of the chromosphere from poles to the equator and the complex velocity distribution in active regions are clearly displayed.

The CIM Level 1 images are produced from Level 0 data via dark- and flat-field correction, and coordinate transformation. The flat field is obtained by using the normal KLL method \cite{kuhn}, which, at first, records multiple images of the solar disk centered at different positions of the detector by changing the satellite pointing, and then, iterates these images by specific algorithm. The CIM Level 1 data can be used to study the solar activities in the photosphere and to verify the stability of the satellite platform.

\subsection{Data distribution}

The FITS formatted Level 1 science data will be available to the community through the website at SSDC-NJU (\url{https://ssdc.nju.edu.cn}). The CHASE science team provides routines in the IDL and Python languages to read the FITS files. For higher-level products, users are encouraged to develop new routines. One may contact the CHASE science team for more information and contributions to the software.

\section{Conclusion}\label{sec:5}

The CHASE mission aims to test a newly developed satellite platform and to perform solar spectroscopic observations. The HIS instrument onboard the CHASE mission acquires, for the first time in space, the full-Sun or region-of-interest H$\alpha$ spectroscopic images with high spectral and temporal resolutions. Since the launch of the CHASE mission on October 14, 2021 and the first-light imaging on October 24, 2021, the routine observations have begun. The on-orbit performance of the satellite platform and the HIS instrument is excellent and meets the pre-launch expectations. Calibration work is still being developed, but the calibrated data will be available in the near future to the public. The CHASE/HIS is expected to advance our understanding of the dynamics of solar activity in the lower atmosphere. Combining multi-wavelength observations from both ground- and space-based solar telescopes, the CHASE mission provides critical data sets for the research of solar and stellar physics.

%%%%%%%%%%%%%%%%%%%%%%%%%%%%%%%%%%%%%%%%%%%%%%%%%%%%%%%
%%% Acknowledgements. ??§Ý
%%%%%%%%%%%%%%%%%%%%%%%%%%%%%%%%%%%%%%%%%%%%%%%%%%%%%%%
\Acknowledgements{The CHASE mission is supported by China National Space Administration (CNSA). The efforts to develop a solar space mission such as CHASE require a huge amount of skillful and cross-field collaborations. We would like to thank the team from Shanghai Institute of Satellite Engineering who built the satellite, the team from Changchun Institute of Optics, Fine Mechanics and Physics who designed the scientific payload, and the team from China Center for Resources Satellite Date and Application who provided the service of data transmission. Nanjing University is reponsible for the science and application system. We would also like to thank many individuals who gave valuable comments and suggestions in the development of the CHASE mission: Zhi Xu (Yunnan Observatory, China), Weiqun Gan (Purple Mountain Observatory, China), Jean-Claude Vial (Institut d'Astrophysique Spatiale, France), Weijun Mao and Junping Zhang (Nanjing Insititute of Astronomical Optics and Technology, China).}

%%%%%%%%%%%%%%%%%%%%%%%%%%%%%%%%%%%%%%%%%%%%%%%%%%%%%%%
%%% Conflict of interest. ????????????
%%%%%%%%%%%%%%%%%%%%%%%%%%%%%%%%%%%%%%%%%%%%%%%%%%%%%%%

%%%%%%%%%%%%%%%%%%%%%%%%%%%%%%%%%%%%%%%%%%%%%%%%%%%%%%%
%%% Supplements. ????????, ????
%%%%%%%%%%%%%%%%%%%%%%%%%%%%%%%%%%%%%%%%%%%%%%%%%%%%%%%
%\Supplements{}

%%%%%%%%%%%%%%%%%%%%%%%%%%%%%%%%%%%%%%%%%%%%%%%%%%%%%%%
%%% Reference section. ?¦Ï?????
%%% citation in the content using "some words~\cite{1,2}".
%%% ~ is needed to make the reference number is on the same line with the word before it.
%%%%%%%%%%%%%%%%%%%%%%%%%%%%%%%%%%%%%%%%%%%%%%%%%%%%%%%

%%%%%%%%%%%%%%%%%%%%%%%%%%%%%%%%%%%%%%%%%%%%%%%%%%%%%%%
%%% Appendix sections. ??????, ????
%%%%%%%%%%%%%%%%%%%%%%%%%%%%%%%%%%%%%%%%%%%%%%%%%%%%%%%

\end{multicols}

\begin{thebibliography}{}
	
\bibitem{hale}  G. E. Hale, 1908, Contributions from the Mount Wilson Observatory, \textit{Astrophys. J.} \textbf{27}, 219.

\bibitem{fang1} C. Fang, P. F. Chen, Z. Li, M. D. Ding, Y. Dai, X. Y. Zhang, et al., 2013, \textit{Research in Astron. Astrophys.} \textbf{13}, 1509.

\bibitem{liu} Z. Liu, J. Xu, B. Z. Gu, S. Wang, J. Q. You, L. X. Shen, et al., 2014, \textit{Research in Astron. Astrophys.} \textbf{14}, 705.

\bibitem{potzi} W. P\"{o}tzi, A. M. Veronig, M. Temmer, D. J. Baumgartner, H. Freislich, and H. Strutzmann, 2016, \textit{Solar Phys.} \textbf{291}, 3103.

\bibitem{fang2} C. Fang, B. Z. Gu, X. Y. Yuan, M. D. Ding, P. F. Chen, Z. G. Dai, et al., 2019, \textit{Sci. China-Phys. Mech. Astron.} \textbf{49}, 059603.

\bibitem{rao1} C. H. Rao, N. T. Gu, X. J. Rao, C. Li, L. Q. Zhang, J. L. Huang, et al., 2020, \textit{Sci. China-Phys. Mech. Astron.} \textbf{63}, 109631.

\bibitem{yan} X. L. Yan, Z. Liu, J. Zhang, and Z. Xu, 2020, \textit{Sci. China Tech. Sci.} \textbf{63}, 1656.

\bibitem{ichimoto} K. Ichimoto, T. T. Ishii, K. Otsuji, G. Kimura, Y. Nakatani, N. Kaneda, et al., 2017, \textit{Solar Phys.} \textbf{292}, 63.

\bibitem{berezin} I. A. Berezin, A. G. Tlatov, and N. N. Skorbezh, 2021, \textit{Geomagnetism and Aeronomy} \textbf{61}, 1075.

\bibitem{rao2} C. H. Rao, L. Zhu, X. J. Rao, L. Q. Zhang, H. Bao, L. Kong, et al., 2016, \textit{Astrophys. J.} \textbf{833}, 210.

\bibitem{zhangp} P. Zhang, X. Hu, Q. F. Lu, A. J. Zhu, M. Y. Lin, L. Sun, et al., 2022, \textit{Adv. Atoms. Sci.} \textbf{39}, 1.

\bibitem{gan} W. Q. Gan, C. Zhu, Y. Y. Deng, H. Li, Y. Su, H. Y. Zhang, et al., 2019, \textit{Research in Astron. Astrophys.} \textbf{19}, 156.	

\bibitem{chen1} P. F. Chen, 2018, \textit{Sci. China-Phys. Mech. Astron.} \textbf{61}, 109631.

\bibitem{zhang1} W. Zhang, W. Q. Cheng, W. You, X. Chen, J. Zhang, C. Li, and C. Fang, 2022, \textit{Sci. China-Phys. Mech. Astron.} \textbf{XX}, XXXXX.

\bibitem{lic} C. Li, C. Fang, Z. Li, M. D. Ding, P. F. Chen, Z. Chen, et al., 2019, \textit{Research in Astron. Astrophys.} \textbf{19}, 165.

\bibitem{chen2} P. F. Chen, A. A. Xu, and M. D. Ding, 2020, \textit{Res. Astron. Astrophys.} \textbf{20}, 166.

\bibitem{zhou} Y. H. Zhou, P. F. Chen, J. Hong, and C. Fang, 2020, \textit{Nat. Astron.} \textbf{4}, 994.

\bibitem{chen3} P. F. Chen, D. E. Innes, and S. K. Solanki, 2008, \textit{Astron. Astrophys.} \textbf{484}, 487.

\bibitem{yang} L. H. Yang, X. L. Yan, T. Li, Z. K. Xue, and Y. Y. Xiang, 2017, \textit{Astrophys. J.} \textbf{838}, 131.

\bibitem{devore} C. R. DeVore, and S. K. Antiochos, 2000, \textit{Astrophys. J.} \textbf{539}, 954.

\bibitem{cheng} X. Cheng, M. D. Ding, and J. Zhang, 2014, \textit{Astrophys. J. Lett.} \textbf{789}, 35.

\bibitem{ouyang} Y. Ouyang, Y. H. Zhou, P. F. Chen, and C. Fang, 2017, \textit{Astrophys. J.} \textbf{835}, 94.

\bibitem{chen4} P. F. Chen, L. K. Harra, and C. Fang, 2014, \textit{Astrophys. J.} \textbf{784}, 50.

\bibitem{casini} R. Casini, S. M. White, and P. G. Judge, 2017, \textit{Space Sci. Rev.} \textbf{210}, 145.

\bibitem{ding} M. D. Ding, C. Fang, and H. S. Yun, 1999, \textit{Astrophys. J.} \textbf{512}, 454.

\bibitem{fang3} C. Fang, Y. H. Tang, Z. Xu, M. D. Ding, and P. F. Chen, 2006, \textit{Astrophys. J.} \textbf{643}, 1325.

\bibitem{wang} H. M. Wang, C. Liu, K. Ahn, Y. Xu, J. Jing, N. Deng, et al., \textit{Nat. Astron.} \textbf{1}, 85.

\bibitem{schroter} E. H. Schr\"{o}ter, 1985, \textit{Solar Phys.} \textbf{100}, 141.

\bibitem{lik} K. J. Li, J. C. Xu, J. L. Xie, and W. Feng, 2020, \textit{Astrophys. J. Lett.} \textbf{905}, 11.

\bibitem{brooks} D. H. Brooks, I. Ugarte-Urra, and H. P. Warren, 2015, \textit{Nat. Commun.} \textbf{6}, 5947.

\bibitem{pesnell} W. D. Pesnell, B. J. Thompson, and P. C. Chamberlin, 2012, \textit{Solar Phys.} \textbf{275}, 3.

\bibitem{pontieu} B. De Pontieu, A. M. Title, J. R. Lemen, J. D. Kushner, D. J. Akin, B. Allard, et al., 2014, \textit{Solar Phys.} \textbf{289}, 2733.

\bibitem{muller} D. M\"{u}ller, O. C. St. Cyr., I. Zouganelis, H. R. Gilbert, R. Mason, T. Nieves-Chinchilla, et al., 2020, \textit{Astron. Astrophys.} \textbf{642}, A1.

\bibitem{schaefer} B. E. Schaefer, 1989, \textit{Astrophys. J.} \textbf{337}, 927.

\bibitem{balona} L. A. Balona, 2015, \textit{Mon. Not. Roy. Astron. Soc.} \textbf{447}, 2714.

\bibitem{odert} P. Odert, M. Leitzinger, A. Hanslmeier, and H. Lammer, 2017, \textit{Mon. Not. Roy. Astron. Soc.} \textbf{472}, 876.

\bibitem{argiroffi} C. Argiroffi, F. Reale, J. J. Drake, A. Ciaravella, P. Testa, R. Bonito, et al., 2019, \textit{Nat. Astron.} \textbf{3}, 742.

\bibitem{Zhang2} W. Zhang, Y. Yang, W. You, H. Yu, F. Q. Li, X. Chen, et al., 2022, \textit{Sci. China-Phys. Mech. Astron.} \textbf{XX}, XXXXXX.

\bibitem{Liu1} Q. Liu, et al., 2022, \textit{Sci. China-Phys. Mech. Astron.} \textbf{XX}, XXXXXX.

\bibitem{chenz} Z. Chen, X. X. Zhang, C. Z. Chen, J. Y. Ren, 2016, \textit{Chinese Journal of Lasers} \textbf{4}, 239.

\bibitem{Qiu1} Y. Qiu, S. H. Rao, C. Li, M. D. Ding, C. Fang, Z. Li, et al., 2022, \textit{Sci. China-Phys. Mech. Astron.} \textbf{XX}, XXXXXX.

\bibitem{kuhn} J. R. Kuhn, H. Lin, and D. Loranz, 1991, \textit{PASJ} \textbf{103}, 1097.

\end{thebibliography}
\end{document}